\documentclass[english]{elsart}
\usepackage[T1]{fontenc}
\usepackage[latin1]{inputenc}
\usepackage{amsmath}
\usepackage{setspace}
\usepackage{amssymb}
\usepackage{color}
\makeatletter
\providecommand{\LyX}{L\kern-.1667em\lower.25em\hbox{Y}\kern-.125emX\@}
\usepackage{setspace}
\usepackage[numbers]{natbib}
\usepackage[dvips]{graphicx}
\usepackage{epic,epsfig}
\usepackage{subfigure,float}
\usepackage{amsmath,amsfonts,amssymb}
\usepackage{inputenc}
\usepackage{natbib}
\usepackage{euscript}
\makeatother

\begin{document}
\begin{frontmatter}

\title{A Model for Social Networks}
\author{Riitta Toivonen\corauthref{cor1}}, 
\ead{rtoivone@lce.hut.fi}
\author{Jukka-Pekka Onnela},
\author{Jari Saram\"{a}ki},
\author{J\"{o}rkki Hyv\"{o}nen}, and
\author{Kimmo Kaski}
\corauth[cor1]{Corresponding author.}
\address{Laboratory of Computational Engineering, Helsinki University
  of Technology, P.O. Box 9203, FIN-02015 HUT, Finland}

\begin{abstract}
Social networks are organized into communities with dense internal
 connections, giving rise to high values of the clustering
 coefficient.  In addition, these networks have been observed to be
 assortative, i.e. highly connected vertices tend to connect to other
 highly connected vertices, and have broad degree distributions.  We
 present a model for an undirected growing network which reproduces
 these characteristics, with the aim of producing efficiently very
 large networks to be used as platforms for studying sociodynamic
 phenomena. The communities arise from a mixture of random attachment
 and implicit preferential attachment. The structural properties of
 the model are studied analytically and numerically, using the
 $k$-clique method for quantifying the communities.
\end{abstract}

\end{frontmatter}

\section{\label{sec:intro} Introduction}

The recent substantial interest in the structural and functional
properties of complex networks (for reviews, see
\cite{Albert:review,Dorogovtsev:evolution, Newman:review}) has been
partially stimulated by attempts to understand the characteristics of
social networks, such as the small-world property and high degree of
clustering~\cite{SW}.  Before this, social networks have been
intensively studied by social scientists~\cite{Milgram67,
Granovetter73,Wasserman} for several decades in order to understand
both local phenomena, such as clique formation and their dynamics, as
well as network-wide processes, such as transmission of information.
Within the framework of complex networks, studies have concentrated on
the structural analysis of various types of social networks, such as
those related to sexual contacts~\cite{LiljerosSexweb}, professional
collaboration~\cite{SW,NewmanCollaboration2001,NewmanScientific} and
Internet dating~\cite{HolmeDating}, as well as models of collective
behaviour and various sociodynamic
phenomena~\cite{ZanetteRumor,KlemmSocial, MorenoRumor}. One feature of
particular interest has been to evaluate and detect community
structure in
networks~\cite{GirvanPNAS,NewmanCommunity,NewmanFast,OverlappingCommunities},
where the developed methodologies have found applications in various
other fields such as systems
biology~\cite{GuimeraCartography}. Communities can, roughly speaking,
be defined as sets of vertices with dense internal connections, such
that the inter-community connections are relatively sparse. In
everyday social life or professional collaborations, people tend to
form communities, the existence of which is a prominent characteristic
of social networks and has far reaching consequences on the processes
taking place on them, such as propagation of information and opinion
formation.

It is evident that theoretical studies of processes and collective
behaviour taking place on social networks would benefit from realistic
social network models. Essential characteristics for social networks
are believed to include assortative
mixing~\cite{NewmanMixingPRL,NewmanDifferent}, high clustering, short
average path lengths, broad degree
distributions~\cite{AmaralPNAS,BogunaSocial,nd}, and the existence of
community structure.  Here, we propose a new model that exhibits all
the above characteristics. So far, different approaches have been
taken to define social network
models~\cite{BogunaSocial,WongSpatial,Jin2001,SecederModel,Davidsen2002,
NewmanAssortative,LiJPhysA2005,Marsili2004}.  To our knowledge, of the
above \cite{BogunaSocial} exhibits community structure, high
clustering and assortativity\footnote{The model presented in
\cite{SecederModel} also exhibits community structure and high
clustering, but weak assortativity, with assortative mixing
coefficients of the order $0.01$.}, but based on visualizations given
in the paper their community structure appears very different from the
proposed model.  Our model belongs to the class of growing network
models, i.e. all edges are generated in connection with new vertices
joining the network. Network growth is governed by two processes:
1)~attachment to random vertices, and 2)~attachment to the
neighbourhood of the random vertices ("getting to know friends of
friends"), giving rise to implicit preferential attachment. These
processes then, under certain conditions, give rise to broad degree
distributions, high clustering coefficients, strong positive
degree-degree correlations and community structure.

This paper is structured as follows: First, we motivate the model
based on real-world observations, followed by description of the
network growth algorithm. Next, we derive approximate expressions for
the degree distribution and clustering spectrum and compare our
theoretical results to simulations. We also present numerical results
for the degree-degree correlations. We then address the issue of
community structure using the $k$-clique
method~\cite{OverlappingCommunities}.  Finally, we conclude with a
brief summary of our results.

\section{\label{sec:unweighted}Model}

\subsection{\label{sec:requirements}Motivation for the model}

Our basic aim has been to develop a model which a) captures the
salient features of real-world social networks, and b) is as simple as
possible, and simple enough to allow approximate analytical
derivations of the fundamental characteristics, although one of the
desired structural characteristics (positive degree-degree
correlations) makes exact derivations rather difficult.  The resulting
network is of interest rather than the growth mechanism.

To satisfy the first criterion, we have set the following requirements
for the main characteristics of networks generated by our model: i)
Due to limited social resources, the degree distribution $p(k)$ should
have a steep tail~\cite{AmaralPNAS}, ii) Average path lengths should
grow slowly with network size, iii) The networks should exhibit high
average clustering, iv) The networks should display positive
degree-degree correlations, i.e. be assortative, v) The networks
should contain communities with dense internal connections.

Requirement i) is based on the observation that many social
interaction networks display power-law-like degree distributions but
may display a cutoff at large
degrees~\cite{NewmanCollaboration2001,NewmanScientific}. In some
cases, degree exponents beyond the commonly expected range
$2<\gamma\leq 3$ have been observed, e.g., in the PGP web of
trust~\cite{BogunaSocial} a power-law like tail with exponent
$\gamma=4$ has been observed. Similar findings have also been made in
a study based on a very large mobile phone call dataset~\cite{nd}.  In
light of these data, we will be satisfied with a model that produces
either steep power laws or a cutoff at high degrees.  In the case of
everyday social networks, common sense tells us that even in very
large networks, no person can have tens of thousands of
acquaintances. Hence, if the degree distribution is to be
asymptotically scale-free $p(k) \propto k^{-\gamma}$, the value of the
exponent $\gamma$ should be above the commonly observed range of
$2<\gamma\leq3$ such that in networks of realistic sizes, $N \geq
10^6$ vertices, the maximum degree is limited\footnote{For networks
with a scale-free tail of the degree distribution, $k_{max}\sim
N^{1/\left(\gamma-1\right)}$.}, $k_{max}\sim 10^2$. As detailed later,
such power-law distributions can be attributed to growth processes
mixing random and preferential attachment.  

Requirement ii), short average path lengths, is a common
characteristic observed in natural networks, including social
networks. Requirements iii) high clustering, iv) assortativity, and v)
existence of communities are also based on existing observations, and
can be attributed to "local" edge formation, i.e. edges formed between
vertices within short distances. The degree of clustering is typically
measured using the average clustering coefficient $\left<c\right>$,
defined as the network average of $c(k)=2E/ k\left(k-1\right)$, where
$E$ is the number of triangles around a vertex of degree $k$ and the
factor $\frac{1}{2}k\left(k-1\right)$ gives the maximum number of such
triangles. A commonly utilized measure of degree-degree correlations
is the average nearest-neighbour degree spectrum $k_{nn}(k)$ - if
$k_{nn}(k)$ has a positive slope, high-degree vertices tend to be
connected to other high-degree vertices, i.e. the vertex degrees in
the network are assortatively mixed (see, e.g.,
Ref.~\cite{PastorSatorrasPRL2001}). For detecting and characterizing
communities, several methods have been proposed~\cite{GirvanPNAS,
NewmanCommunity, NewmanFast,
OverlappingCommunities,GuimeraCartography}.  In social networks, each
individual can be assigned to several communities, and thus we have
chosen to investigate the community structure of our model networks
using a method which allows membership in several
communities~\cite{OverlappingCommunities}.

 To satisfy the second criterion, we have chosen a growing network model,
since this allows using the rate equation
approach~\cite{Barabasi1999,Szabo2003}, and because even very large
networks can be produced using a simple and quick algorithm.  It has
been convincingly argued~\cite{Jin2001} that since the number of
vertices in a social network changes at a very slow rate compared to
edges, a realistic social network model should feature a fixed number
of vertices with a varying number and configuration of edges.
However, as our focus is to merely provide a model generating
substrate networks for future studies of sociodynamic phenomena, the
time scales of which can be viewed to be much shorter than the time
scales of changes in the network structure, a model where the networks
are grown to desired size and then considered static is suitable for
our purposes.

\subsection{\label{sec:algorithm}Model algorithm}

The algorithm consists of two growth processes: 1)~random attachment,
and 2)~implicit preferential attachment resulting from following edges
from the randomly chosen initial contacts.  The local nature of the
second process gives rise to high clustering, assortativity and
community structure. As will be shown below, the degree distribution
is determined by the number of edges generated by the second process
for each random attachment.  The algorithm of the model reads as
follows\footnote{Our network growth mechanism bears some similarity to
the Holme-Kim model, designed to produce scale-free networks with high
clustering~\cite{HKModel}. In the HK model, the networks are grown
with two processes: preferential attachment and triangle formation by
connections to the neighbourhood. However, the structural properties
of networks generated by our model differ considerably from HK model
networks (e.g. in terms of assortativity and community structure).}:

\begin{enumerate}
\item Start with a seed network of $N_0$ vertices.  
\item  \label{item:initialcontacts}
  Pick on average $m_r\geq 1$ random vertices as initial contacts.
\item \label{item:walks} Pick on average $m_s\geq 0$ neighbours of
each initial contact as secondary contacts.
\item \label{item:links} Connect the new vertex to the initial and
secondary contacts.
\item Repeat steps \ref{item:initialcontacts} to \ref{item:links}
until the network has grown to desired size.
\end{enumerate}

\begin{figure}[htb]
\centering
\includegraphics[height=0.25\linewidth]{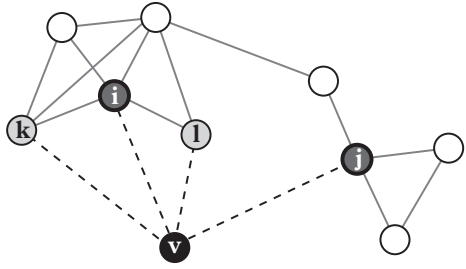} 
  \caption{Growth process of the network. The new vertex $v$ links to
  one or more randomly chosen initial contacts (here $i,j$) and
  possibly to some of their neighbours (here $k,l$). Roughly speaking,
  the neighbourhood connections contribute to the formation of
  communities, while the new vertex acts as a bridge between
  communities if more than one initial contact was chosen. }
  \label{fig:schematic}
\end{figure} 

\begin{figure}[htbp!]
\centering
\includegraphics[width=0.95\linewidth]{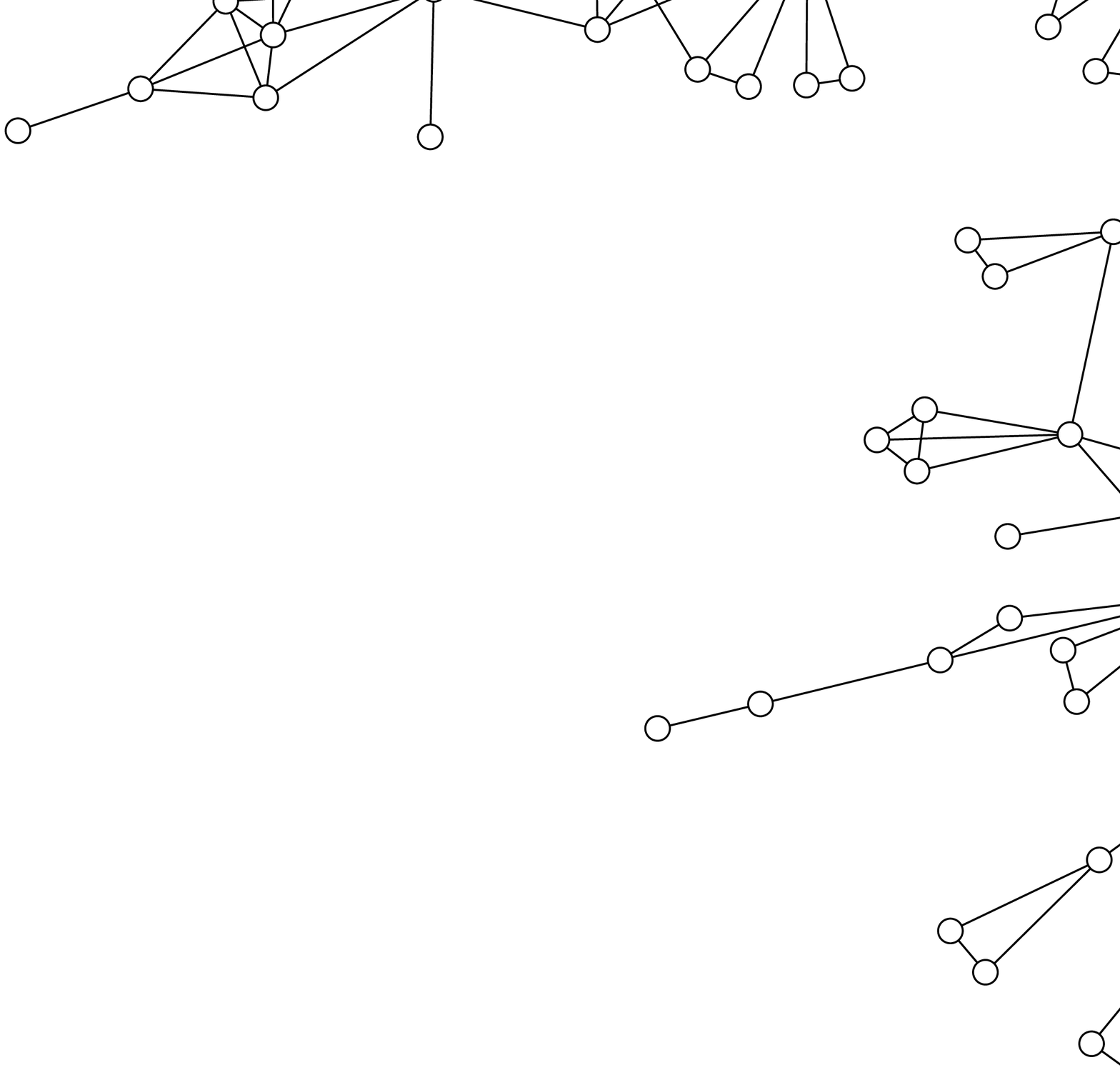} 
\caption{A visualization of a small network with $N=500$ indicates
  strong community structure with communities of various sizes clearly
  visible.  The number of initial contacts is distributed as
  $p(n_{init}\!=\!1)=0.95$, $p(n_{init}\!=\!2)=0.05$, and the number
  of secondary contacts from each initial contact $n_{2nd} \sim
  U[0,3]$ (uniformly distributed between $0$ and $3$).  The network
  was grown from a chain of $30$ vertices. Visualization was done
  using Himmeli~\cite{Himmeli}.}
  \label{fig:comms_Himmeli}
\end{figure}

The analytical calculations detailed in the next section use the
expectation values for $m_r$ and $m_s$. For the implementation, any
non-negative distributions of $m_r$ and $m_s$ can be chosen with these
expectation values. If the distribution for the number of secondary
contacts has a long tail, it will often happen that the number of
attempted secondary contacts is higher than the degree of the initial
contact so that all attempted contacts cannot take place, which will
bias the degree distribution towards smaller degrees. We call this the
\emph{saturation} effect, since it is caused by all the neighbours of
an initial contact being used up, or saturated.  However, for the
distributions of $m_s$ used in this paper the saturation effect does
not seem to have much effect on the degree distribution.

For appreciable community structure to form, it is essential that the
number of links made to the neighbors of an initial contact varies,
instead of always linking to one or all of the neighbors, and that
sometimes more than one initial contact are chosen, to form 'bridges
between communities'.  Here, we use the discrete uniform distributions
$n_{2nd} \sim U[0,k], \; k=1,2,3$ for the number of secondary contacts
$n_{2nd}$, while for the number of initial contacts $n_{init}$ we
usually fix the probabilities to be $p_1=0.95$ for picking one contact
and $p_2=0.05$ for picking two. This results in sparse connectivity
between the communities.  The uniform distributions for $n_{2nd}$ were
chosen for simplicity, but allowing larger $n_{2nd}$ would allow for
larger cliques and stronger communities to form.

\subsection{\label{sec:degree}Vertex degree distribution}

We will use the standard mean-field rate equation
method~\cite{Barabasi1999} to derive an approximative expression for
the vertex degree distribution.  For growing network models mixing
random and preferential attachment, power law degree distributions
$p(k) \sim k^{\gamma}$ with exponents $2 < \gamma < \infty$ have been
derived in e.g. ~\cite{KRGen,JSscalefree,SergeiGen}\footnote{The same
result is found for generalized linear preferential attachment kernels
$\pi_k \propto k+k_0$, where $k_0$ is a constant, since mixing random
and preferential attachment can be recast as preferential attachment
with a shifted kernel.}.  Since in our model the newly added links
always emanate from the new vertex, the lower bound for the degree
exponent is 3; by contrast, if links are allowed to form between
existing vertices in the network, the exponent can also have values
between 2 and 3~(see, e.g., \cite{JSscalefree}).

If no degree correlations were present, choosing a vertex on the other
end of a randomly selected edge would correspond to linear
preferential selection.  In this model network correlations are
present, leading to a bias from pure preferential attachment.
Qualitatively, this can be explained as follows: A low degree vertex
will have on the average low degree neighbors.  Therefore, starting
from a low degree vertex, which are the most numerous in the network,
and proceeding to the neighbourhood, we are more likely to reach low
degree vertices than their proportion in the network would imply.
Hence, the hubs gain fewer links than they would with pure
preferential attachment.  Due to degree-degree correlations, then, the
simulated curves will not closely match the theory, but at high values
of $k$ the theoretical distributions can be viewed as an upper limit
to the average maximum degrees.

We first construct the rate equation which describes how the degree of
a vertex changes on average during one time step of the network growth
process.  The degree of a vertex $v_i$ grows via two processes: 1)~a
new vertex directly links to $v_i$ (the probability of this happening
is $m_r / t$, since there are altogether $\sim t$ vertices at time
$t$, and $m_r$ random initial contacts are picked) 2)~vertex $v_i$ is
selected as a secondary contact.  In the following derivations we
assume that the probability of 2) is linear with respect to vertex
degree, i.e. following a random edge from a randomly selected vertex
gives rise to implicit preferential attachment.  Note that in this
approximation we neglect the effects of correlations between the
degrees of neighbouring vertices.  On average $m_s$ neighbours of the
$m_r$ initial contacts are selected to be secondary contacts. These
two processes lead to the following rate equation for the degree of
vertex $v_i$:
\begin{eqnarray}
\label{eq:degreerate}
\frac{\partial k_i }{\partial t} &=& m_r \Big( \frac{1}{t} +
m_s\frac{k_i}{\sum k} \Big) = \frac{1}{t} \Big( m_r + \frac{m_s}{ 2 (1
+ m_s) } k_i \Big),
\end{eqnarray}
where we substituted $2 m_r (1 + m_s) \,t$ for $\sum k$, based on the
facts that the average initial degree of a vertex is
$k_{init}=m_r(1+m_s)$, and that the contribution of the seed to the
network size can be ignored.  Separating and integrating (from $t_i$
to $t$, and from $k_{init}$ to $k_i$), we get the following time
evolution for the vertex degrees:
\begin{equation}
\label{eq:k_evolution}
k_i(t) = B \Big( \frac{t}{t_i} \Big)^{1/A} - C,
\end{equation}
where $A=2\left(1+m_s\right)/m_s$,   $B=m_r\left(A+1+m_s\right)$,
 and $C = A m_r$. 

From the time evolution of vertex degree $k_i(t)$ we can calculate the
degree distribution $p(k)$ by forming the cumulative distribution
$F(k)$ and differentiating with respect to $k$.  Since in the mean
field approximation the degree $k_i(t)$ of a vertex $v_i$ increases
strictly monotonously from the time $t_i$ the vertex is initially
added to the network, the fraction of vertices whose degree is less
than $k_i(t)$ at time $t$ is equivalent to the fraction of vertices
that were introduced after time $t_i$. Since $t$ is evenly
distributed, this fraction is $(t - t_i)/t$.  These facts lead to the
cumulative distribution
\begin{equation}
\label{eq:k_cumulative}
F( k_i ) = P( \,\tilde{k} \le k_i \,) = 
                   P( \,\tilde{t} \ge t_i \,) 
 = 
\frac{1}{t} \,(\,t - t_i \,). 
\end{equation}

Solving for $t_i = t_i (k_i,t) = B^A\, (k_i + C)^{-A}\, t \,$ from
(\ref{eq:k_evolution}) and inserting it into (\ref{eq:k_cumulative}),
differentiating $F( k_i )$ with respect to $k_i$, and replacing the
notation $k_i$ by $k$ in the resulting equation, we get the
probability density distribution for the degree $k$ as:
\begin{equation}
p(k) =  A B^A (k + C )^{ -2/m_s - \,3}, \label{eq:degreedistribution}
\end{equation}
\noindent where $A$, $B$ and $C$ are as above. Hence, in the limit of
large $k$, the distribution becomes a power law $p(k) \propto
k^{-\gamma}$, with $\gamma = 3 + \frac{2}{m_s}$, $m_s > 0$,
leading to $3<\gamma<\infty$.  
In the model, $\gamma=3$ can never be
reached due to the random component of attachment. When the importance
of the random connection is diminished with respect to the implicit
preferential component by increasing $m_s$, however, the theoretical
degree exponent approaches the limit 3, the value resulting from pure
preferential attachment.

\subsection{\label{sec:clustering}Clustering spectrum}

The dependence of the clustering coefficient on vertex degree can also
be found by the rate equation method \cite{Szabo2003}. Let us examine
how the number of triangles $E_i$ around a vertex $v_i$ changes with
time. The triangles around $v_i$ are mainly generated by two
processes: \mbox{1) Vertex $v_i$} is chosen as one of the initial
contacts with probability $m_r/t$, and the new vertex links to some of
its neighbours (we assume $m_s$ on average, although sometimes this is
limited by the number of neighbours the initial contact has,
i.e. saturation) \mbox{2) The vertex $v_i$} is selected as a secondary
contact, and a triangle is formed between the new vertex, the initial
contact and the secondary contact.  Note that triangles can also be
generated by selecting two neighbouring vertices as the initial
contacts, but in the first approximation the contribution of this is
negligible. These two processes are described by the rate equation
\begin{equation}
\frac{\partial E_i(k_i,t)}{\partial t} =  \frac{m_r m_s}{t} + 
                                     m_r m_s \, \frac{k_i}{\sum k}
= \frac{\partial k_i}{\partial t} + \frac{m_r(m_s\!-\!1)}{t},  
\end{equation}
where the second right hand side is obtained by applying
Eq.~(\ref{eq:degreerate}).  Integrating both sides with respect to $t$,
and using the initial condition \mbox{$E_i(k_{init},t_i) = m_r(1+m_s)$}, we
get the time evolution of triangles around a vertex $v_i$ as
\begin{equation} 
\label{eq:E_evolution}
  E_i(t) = k_i(t) + m_r( m_s-1) \ln \Big(\frac{t}{t_i}\Big) -m_r.  
\end{equation}

\noindent We can now make use the previously found dependence of $k_i$
on $t_i$ for finding $c_i(k)$.  Solving for $\ln
\big(\frac{t}{t_i}\big)$ in terms of $k_i$ from
(\ref{eq:k_evolution}), inserting it into (\ref{eq:E_evolution}) to
get $E_i(k_i)$, and dividing $E_i(k_i)$ by the maximum possible number
of triangles, $k_i(k_i-1)/2$, we arrive at the clustering coefficient:
\begin{eqnarray}  
c_i(k_i) = \frac{2 E_i(k_i)}{k_i(k_i-1)}
&=& 2 \, \frac{k_i + D \ln (k_i + C) - F}{k_i(k_i-1)}, \nonumber \label{Eq:c}
\end{eqnarray}
where $\,C=Am_r$, $\,D=C( m_s -1)$, and $F= D\, \ln B + m_r$. For
large values of degree $k$, the clustering coefficient thus
depends on $k$ as $c(k) \sim 1/k$.

\subsection{Comparison of theory and simulations}

\begin{figure}[htb]
\centering
\includegraphics[width=1.00\linewidth]{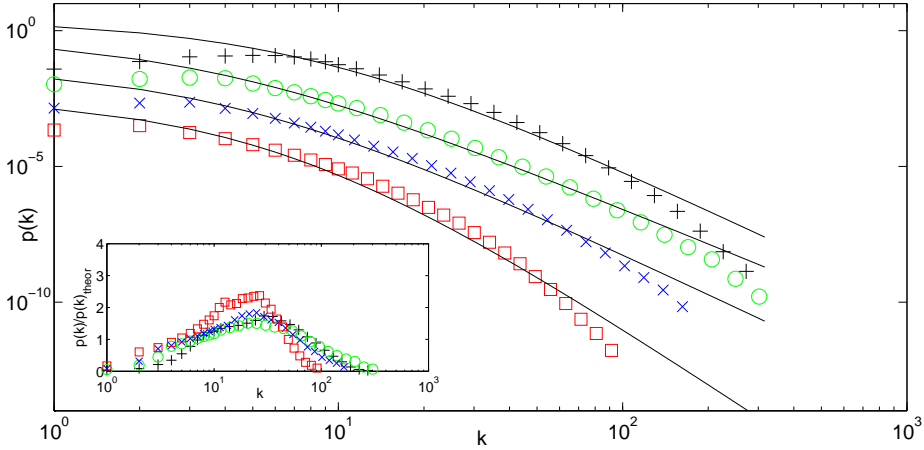} 
   \caption{Degree distributions of simulated networks of size
  $N=10^6$, averaged over 100 runs each. Due to degree-degree
  correlations in the network, linking to the neighbourhood of a
  vertex does not strictly lead to preferential attachment, which
  causes the distributions to fall below the theoretical power laws
  (solid lines) at large $k$.  Curves are vertically translated a
  decade apart for clarity.  Inset: the ratio of simulated values to
  theoretical ones.  Markers correspond to different parameter values:
  ($+$): number of initial contacts $n_{init}$ from the discrete
  uniform distribution $U[1,3]$, number of secondary contacts
  $n_{2nd}$ from $U[0,2]$.  (\textcolor{green}{$\circ$}):
  $p(n_{init}\!=\!1)=0.95$, $p(n_{init}\!=\!2)=0.05$, $n_{2nd} \sim
  U[0,3]$.  (\textcolor{blue}{$\times$}): $p(n_{init}\!=\!1)=0.95$,
  $p(n_{init}\!=\!2)=0.05$, $n_{2nd} \sim U[0,2]$.
  (\textcolor{red}{$\Box$}): $p(n_{init}\!=\!1)=0.95$,
  $p(n_{init}\!=\!2)=0.05$, $n_{2nd} \sim U[0,1]$.}
\label{fig:degrees}
\end{figure}

Fig.~\ref{fig:degrees} displays the degree distributions averaged over
100 runs for networks of size $N=10^6$ for various parametrizations,
together with analytical curves calculated using
Eq.~(\ref{eq:degreedistribution}). The analytical distributions
asymptotically approach power laws with exponents $p(k)\propto
k^{-\gamma}$ (from top to bottom) $\gamma=5$, $4.33$, $5$, and
$7$. The tails of the simulated distributions fall below the
theoretical predictions due to degree correlations, as explained
earlier. The degree-degree correlations were confirmed as the cause of
the deviation by replacing the attachment to secondary contacts by
pure random preferential attachment, after which the simulated and
theoretical slopes matched very closely (not shown). Note that the
parameter values shown here were chosen for simplicity, and they could
be tuned for different qualities.

\begin{figure}[ht]
\centering
 \hspace{-0.2cm}\includegraphics[width=0.8\linewidth]
{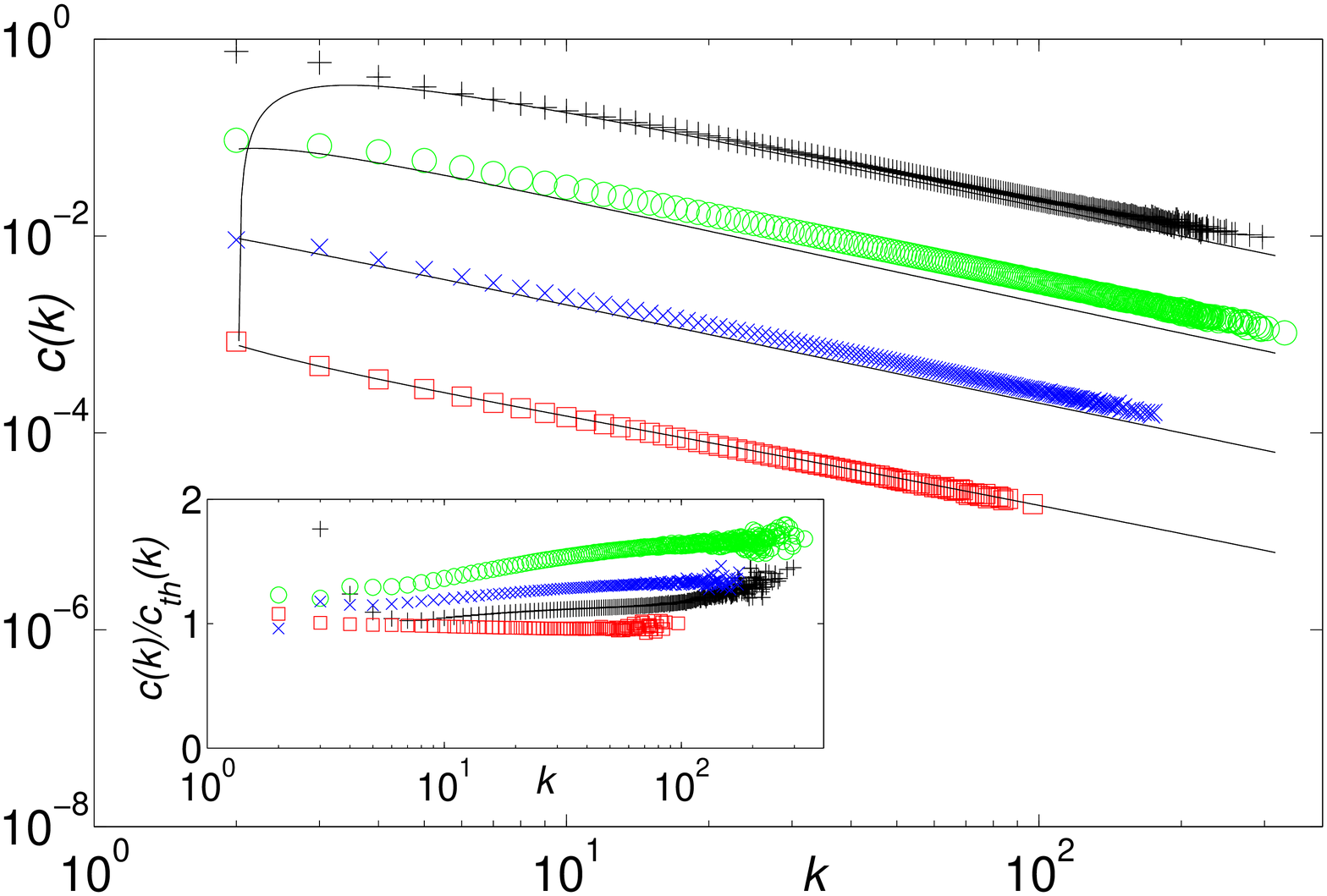} \\
\includegraphics[width=0.50\linewidth,height=0.38\linewidth]{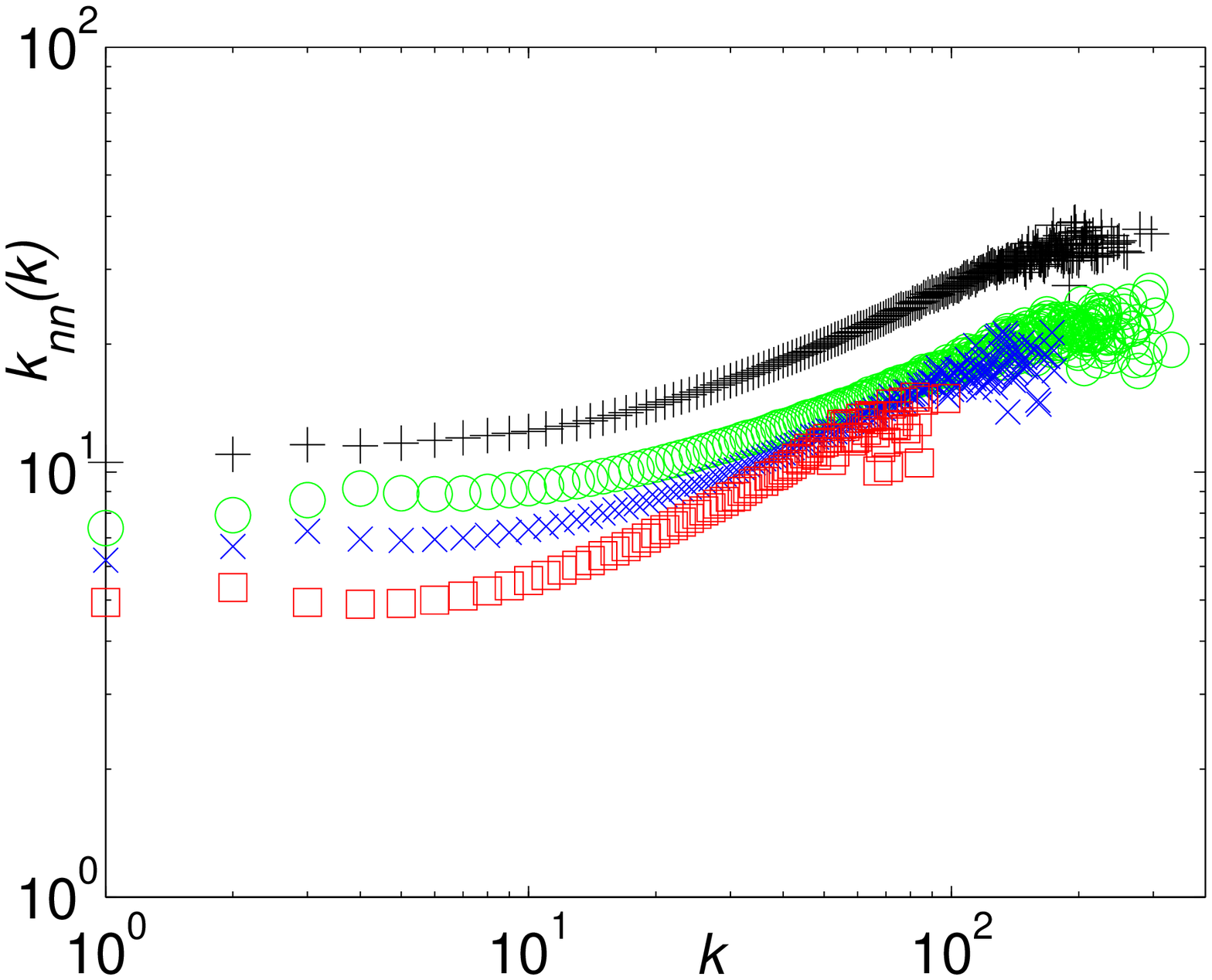} \hspace{-0.6cm}
\includegraphics[width=0.50\linewidth,height=0.38\linewidth]{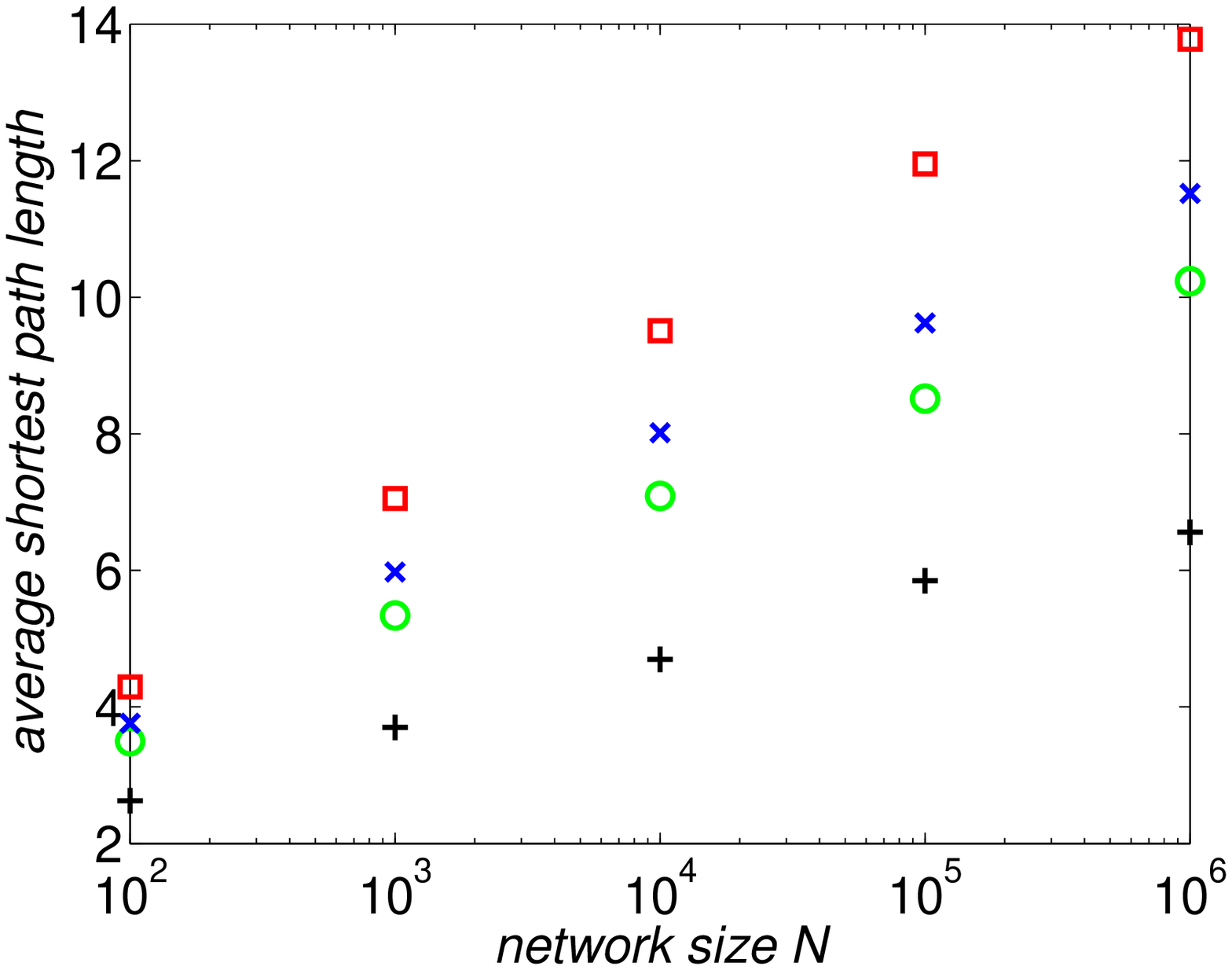} 
  \caption{Top: Clustering coefficient $c(k)$, averaged over 100
  iterations for networks of size $N=10^6$. Predictions for $c(k)$
  (solid lines) agree well with simulated results. Curves are
  vertically translated a decade apart for clarity.  Inset: the ratio
  of simulation results to theory.  Bottom left: Average
  nearest-neighbour degree $k_{nn}(k)$ for the same networks,
  displaying a signature of assortative mixing. Bottom right: average
  shortest path lengths grow logarithmically with network size. ($+$):
  number of initial contacts from $U[1,3]$, secondary contacts from
  $U[0,2]$. Markers correspond to the same parameters as in
  Fig.~\ref{fig:degrees}. }
  \label{fig:c_and_knn}
\end{figure}

\newpage
The top panel of Fig.~\ref{fig:c_and_knn} displays averaged values of
the clustering coefficient $c(k)$ for the same networks, together with
analytical curves calculated using Eq.~(\ref{Eq:c}). We see that the
predictions match the simulated results well, and the $c(k)\sim
1/k$-trend is clearly visible. The corresponding network-averaged
clustering coefficients are (top to bottom) $\left<c\right>=0.30$,
$0.58$, $0.54$ and $0.43$, i.e. the degree of clustering is relatively
high. Of these parameter sets, (\textcolor{green}{$\circ$}) allows the
largest number of links from each initial contact, therefore giving
the largest average clustering. Higher clustering coefficients
could be obtained by increasing the possible number of secondary
contacts.

\newpage
\subsection{\label{sec:knn}Degree-degree correlations and average shortest path lengths}

Next, we investigate the degree-degree correlations of our model
networks.  Social networks are often associated with assortative
mixing~\cite{NewmanMixingPRL} related to vertex degrees,
i.e. high-degree vertices tend to connect to other high-degree
vertices.  This tendency can be formulated in terms of a conditional
probability $P(k^\prime|k)$ that an edge connected to a vertex of
degree $k$ has a vertex of degree $k^\prime$ at its other
end~\cite{PastorSatorrasPRL2001}. A quantity more suitable for
numerical investigations is the average nearest-neighbour degree
$k_{nn}(k)=\sum_{k^\prime} k^\prime P(k^\prime|k)$. If $k_{nn}(k)$ is
an increasing function of $k$, the network is assortatively mixed in
terms of vertex degrees. The bottom left panel in
Fig.~\ref{fig:c_and_knn} shows $k_{nn}(k)$ averaged over 100 networks,
displaying a clear signature of assortative mixing. Another measure of
degree-degree correlations is the assortativity coefficient
$r$~\cite{NewmanAssortative}, which is the Pearson correlation
coefficient of vertex degrees at either end of an edge.  For the model
networks generated with the parameters used in this paper, the
coefficients are ($+$): $0.18$, (\textcolor{green}{$\circ$}): $0.10$,
(\textcolor{blue}{$\times$}): $0.10$, and (\textcolor{red}{$\Box$}):
$0.09$.  For different co-authorship networks, for example, the
assortativity coefficient has been found to range from $0.12$ to
$0.36$~\cite{NewmanMixingPRL}.

Qualitatively, the presence of positive degree-degree correlations can
be attributed to the neighbourhood connections, as well as the high
degree of clustering. Consider a situation where a new vertex attaches
to one initial contact $v_i$ and $m_s$ of its neighbours, so that the
degree of all the vertices in question is increased by one. Hence,
positive correlations are induced between the degrees of $v_i$ and its $m_s$
neighbours. In addition, because of the high clustering, there is a
large probability of connections between the $m_s$ neighbours. This
gives rise to positive degree correlations between the $m_s$ vertices.

It is commonly observed in real life networks that average path
lenghts are short with respect to network size~\cite{SW}.
Together with high clustering, this is called the small world effect.
Typically in model networks, the shortest path lengths are found to
grow logarithmically with network size. This is also the case in 
our model (Fig.~\ref{fig:c_and_knn}, bottom right panel). 

\vspace{20 pt}
\subsection{\label{sec:communities}Community structure}

\begin{figure}[hbp!]
\centering
\includegraphics[width=0.95\linewidth]{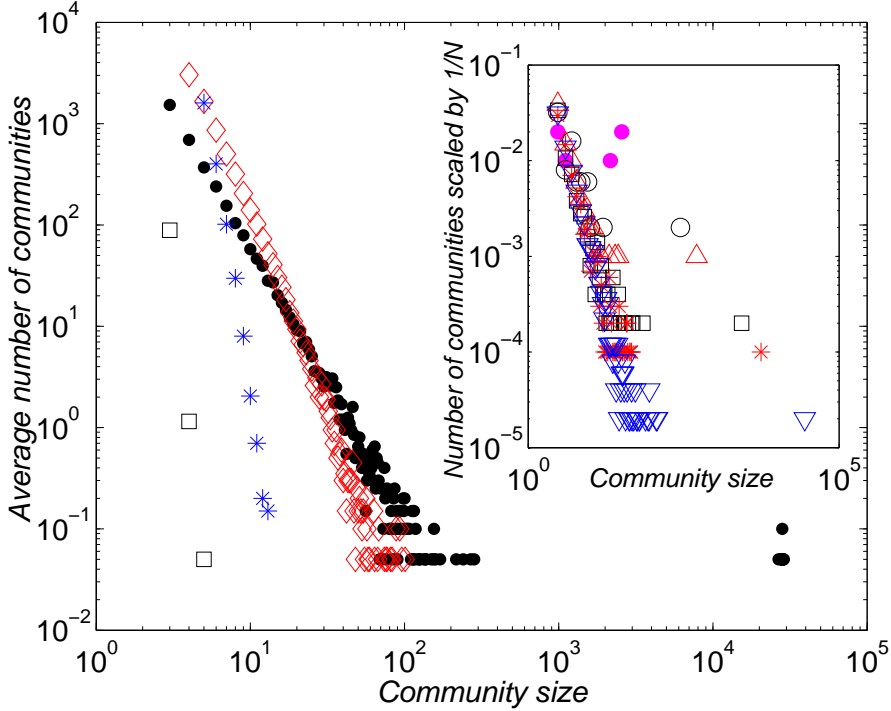}   
\caption{The average number of $k$-clique-communities ( $\bullet$:
  $k\!=\!3$, \textcolor{red}{$\diamond$}: $k\!=\!4$,
  \textcolor{blue}{$\ast$}: $k\!=\!5$) of each size found in our model
  network with $N\!=\!50\, 000,$ number of initial connections
  $p(n_{init}=1)=0.95$, $p(n_{init}=2)=0.05$, and number of secondary
  connections from $U[0,3]$, averaged over 20 networks.  In the case
  of 3-cliques, large communities spanning roughly half the network
  are seen.  The community size distributions are broad, and their
  log-log plots appear power-law-like, although the cumulative
  distributions (not shown) show some deviation. Approximate slopes of
  the log-log plots are $k\!=\!3$: $3$ (excluding the
  supercommunities), $k\!=\!4$: $4$, and $k\!=\!5$: $10$. A very large
  $3$-clique-community spans roughly half of the vertices in any
  network generated with these parameters. In the corresponding
  randomized networks, where edges were shuffled keeping the degree
  distribution intact, there were only a few adjacent triangles, and
  no $4$-cliques at all ($\square$: $3$-clique-communities found in
  the randomized networks). The inset shows the effect of network size
  $N$ on the $3$-clique-community size distribution for
  $N\!=\!100,\,500,\,1000,\,5000,\,10000,\,50000$.  As all data fit on
  the same line when scaled by $1/N$, the network size does not affect
  the slope. Note that different choices of parameters would allow
  larger cliques and larger $k$-clique-communities to form.} 
\label{fig:CommunitySizes}
\end{figure}

The emergence of communities in the networks generated by our model
can be attributed to the effects of the two types of
attachment. Roughly speaking, attachment to the secondary contacts
tends to enlarge existing communities; the new vertex creates
triangles with the initial contact and its nearest neighbours. If the
internal connections within an existing community are dense, the
secondary contacts tend to be members of the same community, and thus
this community grows. On the other hand, new vertices joining the
network may attach to several initial contacts (with our
parametrizations, two or three). If they belong to different
communities, the new vertex assumes the role of a "bridge" between
these. However, no edges are added between the vertices already in the
network. Therefore, the maximum size of a clique, i.e. a fully
connected subgraph, to be found in the network is limited by the
maximum number of edges added per time step. In this model the number
of added edges varies, allowing for fairly large cliques to form while
average vertex degree is kept small.  Visualizations of our model
networks with proper parametrization exhibit clear evidence of
community structure, as shown in Fig.~\ref{fig:comms_Himmeli}.

In order to quantify the community structure, we have utilized the
\emph{k-clique} method of Palla \emph{et
al.}~\cite{OverlappingCommunities,CliquePercolation} and the free
software package CFinder they provide. In this approach, the
definition of communities is based on the observation that a typical
community consists of several fully connected subgraphs (cliques) that
tend to share many of their vertices. Thus, a
$k$-\emph{clique-community} is defined as a union of all $k$-cliques
that can be reached from each other through a series of adjacent
$k$-cliques (where adjacency means sharing $k-1$ vertices). This
definition determines the communities uniquely, and one of its
strengths is that it allows the communities to overlap, i.e. a single
vertex can be a member of several communities. For social networks,
this is especially justified.

We have found that the size distributions of $k$-clique-communities in
our model networks are broad, and appear power-law-like
(Fig.~\ref{fig:CommunitySizes}).  The slopes of the log-log plots were
seen not to depend on the network size $N$.  In the case of 3-cliques,
a very large community spans roughly half of the vertices in any
network generated with these parameters.  Similar large 3-cliques can
be observed in many other networks with communities as well, e.g. in
the datasets provided with the CFinder package: a snapshot of the
co-authorship network of the Los Alamos e-print archives, where $54\%$
of the roughly $30\,000$ vertices belong to the largest
$3$-clique-community; in the word association network of the South
Florida Free Association norms ($67\%$), and in the protein-protein
interaction network of the \emph{Saccharomyces cerevisiae} ($17\%$).
The requirements for a $3$-clique-community are not very strict, and
it is not surprising that one community can span most of the network.
With these choices of parameters, no such supercommunities arise with
$k>3$. 

Comparison of the resulting community size distributions with
randomized networks, where the edges of the networks were scrambled
keeping the degree distributions intact, makes it evident that
community structure is present in the model networks
(Fig.~\ref{fig:CommunitySizes}).  Community sizes depend on i) how the
communities are defined and detected, as different methods divide the
networks into differently sized communities, and ii) what type of
social networks are investigated, as different types of networks can
be expected to display different community structures. Although
analysis of the community structure of empirical social networks is a
relevant question, we will leave it for future work. We attempt to
provide a generic model that can be tuned for desired qualities.

\section{Summary}

In this paper we have developed a model which produces very
efficiently networks resembling real social networks in that they have
assortative degree correlations, high clustering, short average path
lengths, broad degree distributions and prominent community structure.
The model is based on network growth by two processes: attachment to
random vertices and attachment to their neighbourhood.  Theoretical
approximations for the degree distribution and clustering spectrum
have been derived and compared with simulation results.  The observed
deviations can be attributed to degree correlations.  Visualizations
of the networks and quantitative analysis show significant community
structure. In terms of communities defined using the $k$-clique
method, the analyzed community size distributions display
power-law-like tails.  These types of features are also present in
many real-life networks, making the model well suited for simulating
dynamic phenomena on social networks.

\textbf{Acknowledgments}

The authors would like to thank J\'{a}nos Kert\'{e}sz, Tapio Heimo and
Jussi Kumpula for useful discussions. This work has been supported by
the Academy of Finland, project no.~1169043 (the Finnish Center of
Excellence program 2000-2005).

\bibliographystyle{elsart-num}
\bibliography{toivonen}

\end{document}